# The He atom revisited

S Datta[1,2],

[1] Department of Physics, The Icfai Foundation for Higher Education, Hyderabad- 501203, India

J M Rejcek[2] J. L. Fry[2],

[2] Department of Physics, The University of Texas at Arlington, Arlington, Texas 76019

In the Feynman-Kac[1] path integral approach the eigenvalues of a quantum system can be computed using Wiener measure which uses Brownian particle motion. In our previous work[2-3] on such systems we have observed that the Wiener process numerically converges slowly for dimensions greater than two because almost all trajectories will escape to infinity[4]. One can speed up this process by using a Generalized Feynman-Kac (GFK) method[5] in which the new measure associated with the trial function is stationary, so that the convergence rate becomes much faster. We thus achieve an example of 'Importance Sampling' and in the present work we apply it to the Feynman-Kac(FK) path integrals for the ground and first few excited state energies for He to speed up the convergence rate. We calculate the path integrals using space averaging rather than the time averaging as done in the past. The best previous calculations from Variational computations report precisions of $10^{-16}$ Hartrees, whereas in most cases our path integral results obtained for the ground and first excited states of He are lower than these results by about $10^{-6}$ Hartrees or more.

**Introduction**

Helium, unlike hydrogen, is the simplest many-body system which cannot be solved exactly. Further, except for the first excited state, the nodal structure of the electron states is not even known exactly. In this paper we calculate the ground (or nodeless) state and the first excited state whose nodal structure is known exactly. To calculate the $P_z$ state of helium for which we use the necessary nodal conditions [6]. Even in this simplest case for He it is not possible to solve for the eigenstate exactly by analytic methods. The only identified option is to resort to some numerical methods. These include the conventional Variational method, Variational Monte Carlo method, Green's function and Path integral Monte Carlo method. The Variational Monte Carlo method uses Monte Carlo integration and analytic trial functions to compute the eigenvalue and eigenvector estimates of a quantum mechanical system. Green's function Monte Carlo solves the time independent Schrödinger Equation by identifying it with the diffusion equation and simulating this as a random walk with branching. In contrast, the Path Integral Monte Carlo method simulates the exact solution of the Schrödinger Equation using the more mathematically rigorous Feynman-Kac[1] algorithm. Caffarel et al[5] used the path integral technique to study the simple atomic systems and achieved chemical accuracy(2-3 significant figures) in their calculations. By contrast Variational energies are reported in the literature[7,8] with many more significant figures. To date, these elaborate calculations set the non-relativistic benchmarks for energies even though they are not exact. Another Variational work of note is the work by Goldman[9] in which he calculated the ground state energy for He very accurately using a very small basis set. He constructed the basis set that effectively deals with the intrinsic singularity of the Coulomb potential at $r_1 = r_2$, and as a consequence the rate of convergence involved in the calculation increased substantially for the first excited state. But for the ground state, Goldman's trial function does not provide us a good result. We are carrying out further work to understand the reason for it.

**Path Integral Theory**: Cafferel et al[5] pointed out that in the Feynman-Kac method the process is non-ergodic in nature, which makes the procedure inefficient and computationally more expensive. They introduced importance sampling in the ordinary Feynman-Kac path integration by using a reference function $\varphi_0^{(0)}$ which made the associated diffusion process ergodic.. However, the averaging in time they employed to take advantage of the ergodicity of the process made the algorithm serial in nature. For larger systems one needs to apply both importance sampling and parallel programming to make the process numerically more feasible. The goal of the present paper is to implement importance sampling in the FK method in such a way that the algorithm can be essentially parallel in nature and at the same time as rigorous as GFK and as elaborate as Variational Calculations(non-Monte Carlo).

Even though FK formalism provides the basis for rigorous and accurate calculations of ground and excited state properties of many particle systems, it suffers from a slow convergence rate due to the fact that the underlying diffusion process, Brownian motion(Wiener process), is non-recurrent. Specifically, in dimensions higher than two the trajectories of the process escape to infinity with probability one[4]. More precisely, for

$d \geq 3, P\left[\lim_{t \to \infty} R_t = \infty\right] = 1$ (where $R_t$ =Bessel process is the distance of the Brownian particle from the origin) which implies $\lim_{t \to \infty} P(X(t) \in B) = 0$ for any bounded subset B of the d dimensional Euclidean space $R^d$. As a result the sampling within the quantum mechanical region of interaction occurs only during a small fraction of the total simulation time and the rate of the convergence becomes prohibitively slow.

The other path integral method mentioned above, Generalized Feynman-Kac(GFK), was initiated by Soto and Claverie[10] and was subsequently extended to the Full Generalized Feynman-Kac method by Caffarel and Claverie[2]. These procedures can be considered as an application of importance sampling to FK along with the transformation of the Wiener process into Ornstein-Uhlenbeck process, having a distribution determined by both the diffusion and drift terms. The GFK procedure is numerically more feasible because the corresponding distribution exists, i.e., $\lim_{t \to \infty} P(Y(t)) \in B) = \int_B \varphi_0^{(0)2}(x)dx$. The Ornstein-Uhlenbeck process $Y(t)$ has a stationary distribution which makes it highly localized and therefore the convergence rate becomes much faster.

The main objective of this paper is to provide a numerically feasible method for the calculation of the ground and excited states of a many body system which is simple, fast and accurate. For GFK we need a trial function $\varphi_0^{(0)}$, and accuracy in the calculation after a given number of paths depends on the symmetry associated with the trial function and how well it represents the true wave function for the eigenstate desired. In this work we take the Variational Monte Carlo trial functions of Alexander et al [11] and Goldman's CI functions [9] for the ground and first excited state of helium. For the $P_z$ excited state we construct a trial function according to our new nodal condition. These functions give the ground and excited state energies accurate to a few tens of micro-Hartrees.

To develop the GFK solution, we first consider the following Cauchy problem which can be viewed as the time-dependent Schrödinger Equation for purely imaginary time:

$$\frac{\partial u(x,t)}{\partial t} = [\frac{\Delta}{2} - V(x)]u(x,t)$$

with $x \in R^d$ and $u(x,0) = f(x)$ (1)

The FK solution to the above equation is

$$u(x,t) = E_x \left\{ e^{-\int_0^t V(X(s))ds} f(X(t)) \right\}$$ (2)

for $V \in K_\nu$, the Kato class of potentials[12]

A direct benefit of having the above representation is to recover the lowest energy eigenvalue of the Hamiltonian $H = -\frac{\Delta}{2} + V$ for a given symmetry by applying the large deviation principle of Donsker and Varadhan[13]

$$\lambda_1 = \lim_{t \to \infty} -\frac{1}{t} \ln E_x \left[ e^{-\int_0^t V(X(s))ds} f(X(t)) \right]$$ (3)

where $X(t)$ is the standard Wiener process and $E$ is the average over the exponentials with respect to Brownian motion trajectories.

In GFK, we generate a diffusion process with the aid of a twice differentiable non negative function $\varphi_0^{(0)}$ by considering the following Hamiltonian:

$$H_0 = -\frac{\Delta}{2} + V_0$$

$$V_0 = \lambda_0^{(0)} + \frac{\Delta \varphi_0^{(0)}}{\varphi_0^{(0)}}$$

where $\varphi_0^{(0)}$ is the twice differentiable non-negative function and $\lambda_0^{(0)}$ is the energy of $\varphi_0^{(0)}$.

The primary reason for introducing $H_0$ with a localized diffusion process is that it possesses a stationary distribution, unlike the non-localized Brownian motion process that escapes to infinity. In this way one achieves the importance sampling goal in the actual numerical computation, thereby reducing the computing time considerably.

Now let us decompose the Hamiltonian of the quantum mechanical problem into two parts:

$$H = H_0 + V_P$$

$$V_P = V - V_0 = V - \lambda_0^{(0)} - \frac{\Delta \varphi_0^{(0)}}{\varphi_0^{(0)}}$$

The expression for energy in the Generalized Feynman-Kac procedure can now be written in terms of the potential $V_P$ as

$$\lambda_1 = \lambda_0^{(0)} - \lim_{t \to \infty} \frac{1}{t} \ln E_x \left[ \exp\left(-\int_0^t V_P(Y(s))ds\right) \right] \tag{4}$$

where, the diffusion $Y(t)$ solves the following stochastic differential equation[14]

$$dY(t) = \frac{\nabla \varphi_0^{(0)}(Y(t))}{\varphi_0^{(0)}(Y(t))} dt + dX(t)$$

The formalism given above is valid for any arbitrary dimensions d(for a system of N particles in three dimensions $d = 3N$). Generalizations of the class of potential functions for which Eqns 2 and 3 are valid are given by Simon [12] and include most physically interesting potentials, positive or negative, including, in particular, potentials with $1/x$ singularities. It can be argued that the function determined by Eq(3) will be the one with the lowest energy of all possible functions independent of symmetry. Although other interpretations are interesting, the simplest is that the Brownian motion distribution is just a useful mathematical construction which allows one to extract the other physically relevant quantities like density, mean square displacement along with the ground and the excited state energy of a quantum mechanical system. In the numerical implementation of the Eq(2), the 3N dimensional Brownian motion is replaced by 3N independent, properly scaled one dimensional random walks as follows. For a given time t and integer n and we define [15] the vector in $R^{3N}$.

$$W(l) \equiv W(t, n, l) = (w_1^1(t, n, l), w_2^1(t, n, l), w_3^1(t, n, l), \ldots \ldots w_1^N(t, n, l), w_2^N(t, n, l), w_3^N(t, n, l)$$

where $w_j^i(t, n, l) = \sum_{k=1}^{l} \frac{\epsilon_{jk}^i}{\sqrt{n}}$

with $w_j^i(t, n, l) = 0$ for $i = 1, 2, \ldots \ldots N; j = 1, 2, 3$ and $l = 1, 2, \ldots nt$. Here $\epsilon$ is chosen independently and randomly with probability $P$ for all $i, j, k$ such that $P = (\epsilon_{jk}^i = 1) = P(\epsilon_{jk}^i = -1) = \frac{1}{2}$. It is known by an invariance principle [16] that for every $v$ and $W(l)$ defined above,

$$\lim_{n \to \infty} P(\frac{1}{n} \sum_{l=1}^{nt} V(W(l))) \leq v$$

$$= P(\int_0^t V(X(s))ds \leq v$$

Consequently for large n,

$$P[exp(-\frac{1}{n}(\int_0^t V(X(s))ds) \leq v]$$

$$\approx P[exp(-\frac{1}{n} \sum_{l=1}^{nt} V(W(l))) \leq v]$$

By generating $N_{rep}$ independent replications $Z_1, Z_2 \ldots \ldots Z_{N_{rep}}$ of $Z_m = exp(-\frac{1}{n} \sum_{l=1}^{nt} V(W(l)))$ and using the law of large numbers, $(Z_1 + Z_2 + \cdots + Z_{N_{rep}})/N_{rep} = Z(t)$ is an approximation to Eq(2).

Here $W^m(l), m = 1,2 \ldots N_{rep}$ denotes the $m^{th}$ realization of $W(l)$ out of $N_{rep}$ independently run simulations. In the limit of large t and $N_{rep}$ this approximation approaches an equality and forms the basis of a computational scheme for the solution of a many particle system with a prescribed symmetry.

Vartiational Quantum Monte Carlo (VQMC) methods can provide very good trial functions $\varphi_0^{(0)}$ for path integral calculations. In VQMC, the form of the trial function $\varphi_0^{(0)}$ is adjusted so as to minimize the variance of the energy. We have chosen VQMC trial functions[11]) and CI trial functions[9] for the ground and triplet S states of helium. For the $P_z$ state of helium we have used the nodal conditions from our paper[6] and have taken the Variational Monte Carlo energy derived from the trial function as the trial energy in GFK calculations.

Once the trial function is determined, the energy of He is calculated using the discretized version of formula (4). Instead of running a single path for a large time(time average) we calculated energies by averaging over a large number of paths for a large time because we need to calculate the variance using a standard statistical method. Another reason for using the space average is that one can use an algorithm which is essentially parallel. Moreover, our derivation of the Generalized Feynman-Kac formula does not require the trial functions to be square integrable. Any twice differentiable non-negative function can serve our purpose, as opposed to only square integrable functions used in literature.

We simulate Eq(4) by using Bernoulli trials. The Bernoulli random variables are generated via transformations of a uniform r.v. on [0,1][17]. In our computations we used a random number generator called Congruential Generator[18] defined by $S_i = aX_{i-1} + c) \bmod M$ where M is a modulus. We have simulated the $\varepsilon_i$'s by using the generator called 'Superduper', $S_i = (S_{i-1} 69069 + 1) \bmod 2^{32}$ with the initial seed $S_0$ and $\varepsilon_i = 1$ if $S_i > 2^{31}$ and $\varepsilon_i = -1$ if $S_i \leq 2^{31}$. Parameters of the path integral calculations are shown in the table. The stepsize $\Delta x = \frac{1}{scale}$ is in atomic units and was made uniform in space for each atom. The Brownian motion times ranged from 8 to 48 atomic units of time Eq(4). We then extrapolated the value of energy using the monotonicity of $u(x,t)$

The numbers shown in the parenthesis of our results are the errors resulting from the extrapolation procedure. The error bars in the figures reflect the statistical uncertainties involved in the estimation of the path integrals. The other possible sources of error include finite time, finite step size, proximity of electrons to the nucleus, choice of random number generators and the nodal structure of the trial wave functions. Our experience with the step size in this case indicates that if the trial wave functions are very smooth functions of radial variables the calculation is almost step-size independent. While extrapolating our path integral solutions we take extra care to make sure that we have attained the asymptotic limit(by looking at the path integral values and the associated statistical uncertainties) which thereby eliminates the error associated with the finite time. We choose random numbers with a cycle $2^{32}$ such that the chance of recycling is substantially small. We choose our criterion precisely (the numerical factor) in such a way that electrons landing on each other or on the nucleus is ruled out. Finally, since the ground state is nodeless and for the first excited state the nodal structure is known precisely, an exact symmetry can be incorporated in the trial function. For the $P_z$ state of He we incorporate the necessary condition required or that symmetry as we derive it in [6].

**Comparison of the different trial functions:**
Now, we give an account of different trial functions we employed and compare them qualitatively.

1 For the ground and first excited state of He we use trial functions consisting of a fully symmetric/anti-symmetric exponentiated pade. These have been constructed by Alexander et al [8] and can be described as follows:

$$\psi_T^G = (1 + P_{12}) \exp\left(\frac{\sum_{k=0} a_k r_1^n r_2^l r_{12}^m}{\sum_{k=0} b_k r_1^n r_2^l r_{12}^m} - \alpha r_1 - \beta r_2\right)$$

With 20 adjustable parameters $\psi_T^G$ gives a value of -2.9037243(4) when evaluated with 1024000 configuration points.

2. $\psi_T^E = (1 - P_{12}) \exp\left(\dfrac{\sum_{k=0} a_k r_1^n r_2^l r_{12}^m}{\sum_{k=0} b_k r_1^n r_2^l r_{12}^m} - \alpha r_1 - \beta r_2\right)$

With 20 adjustable parameters $\psi_T^E$ gives a value of -2.175228(1) when evaluated with 1024000 configuration points.

3 The trial function in [5]

$\psi(r_1, r_2) = (r_0 - r_1) e^{-\alpha_1 r_1 - \alpha_2 r_2} - (r_0 - r_2) e^{-\alpha_2 r_1 - \alpha_1 r_2}$ with $r_0 = 1, \alpha_1 = 1$ and $\alpha_2 = 2$

This wave function is used to calculate the energy for the first excited state of Helium. This trial function satisfies the electron-nucleus cusp condition. The equation of the nodal surface is given by $r_1 = r_2$. Figures 1 and 2 show the plot of the trial function and the nodal surface respectively.

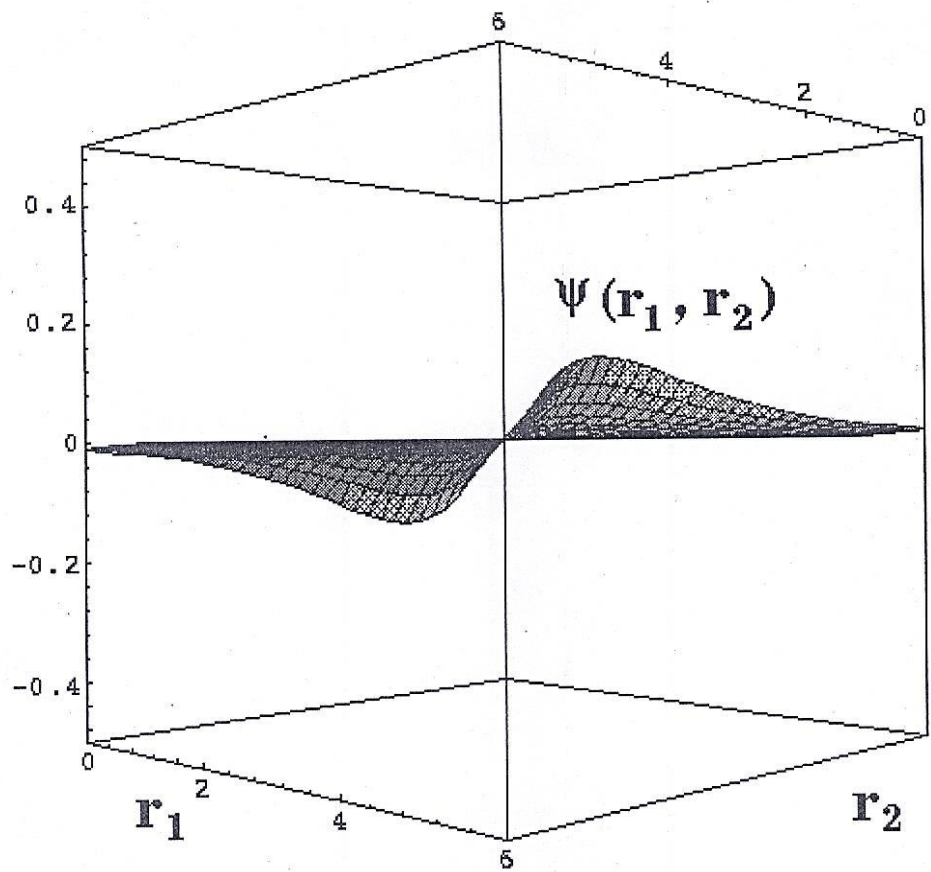

Figure 1   The plot of the wave function 3

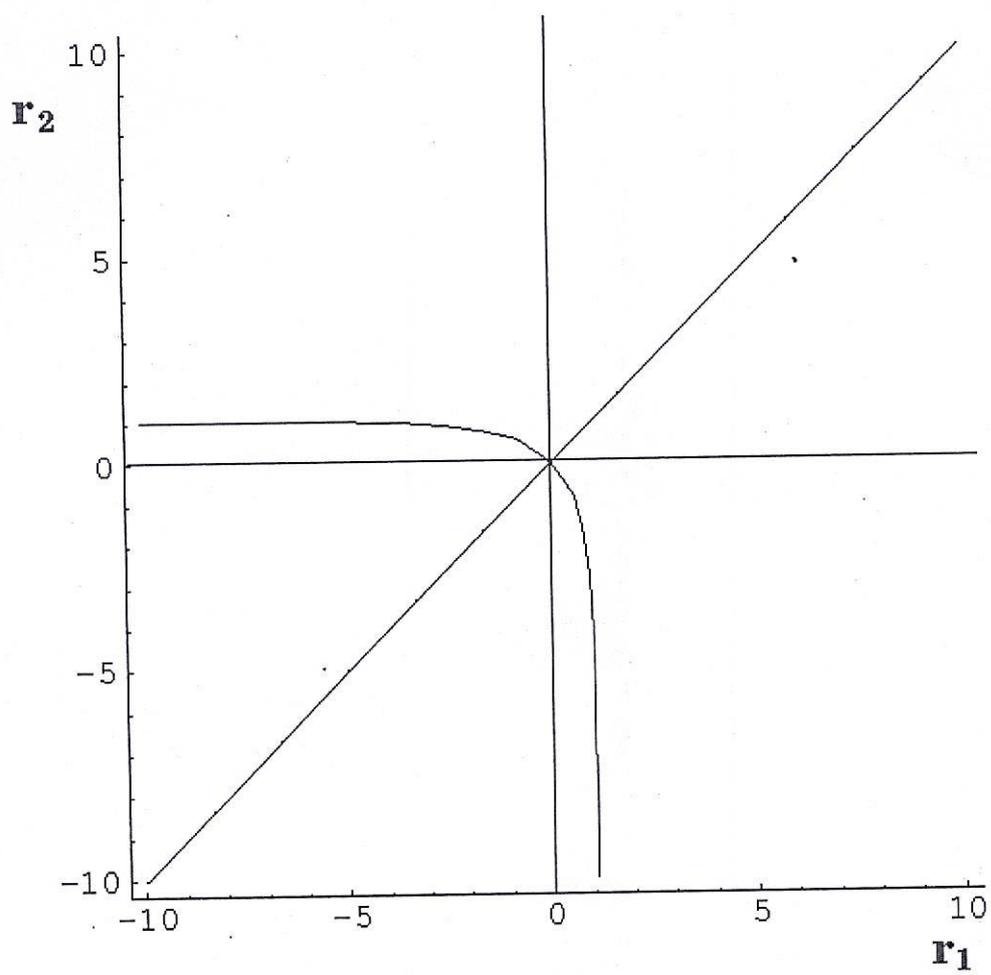

Figure 2  The plot of the nodal surface of the wave function 3

4. The above wave function with $r_0 = 1, \alpha_1 = 0.67180691$ and $\alpha_2 = 2.00411836$.

This wavefunction is used to calculate the energy for the first excited state of the helium atom. It does not satisfy any cusp conditions. Its nodal structure is given by $r_1 = r_2$

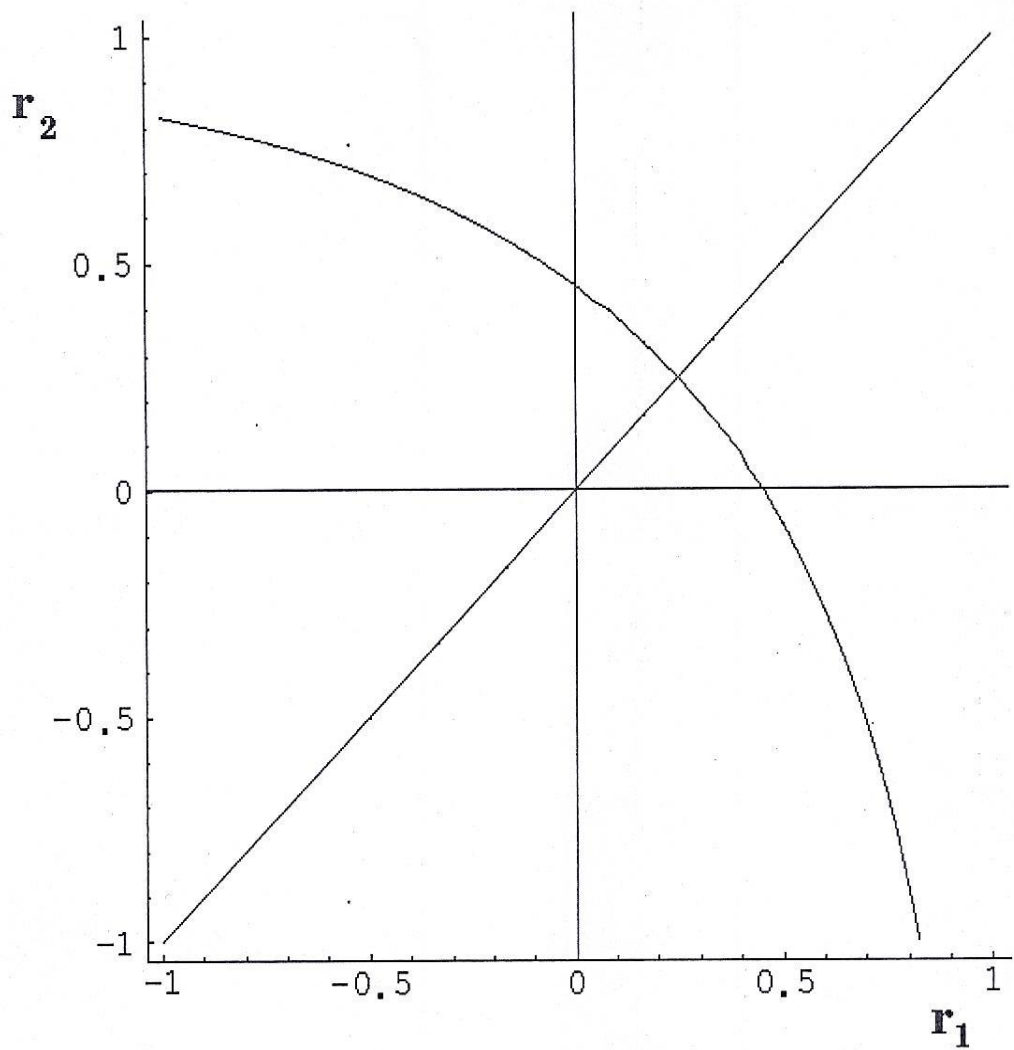

Figure 3  The plot of the nodal surface of the wave function 4

5. The wave function 3 with $r_0 = 0.73351723, \alpha_1 = 0.636748$ and $\alpha_2 = 2.002777$.

This wave function is also used to calculate the energy for the first excited state of helium. It does not satisfy any cusp conditions. Its nodal structure is given by $r_1 = r_2$

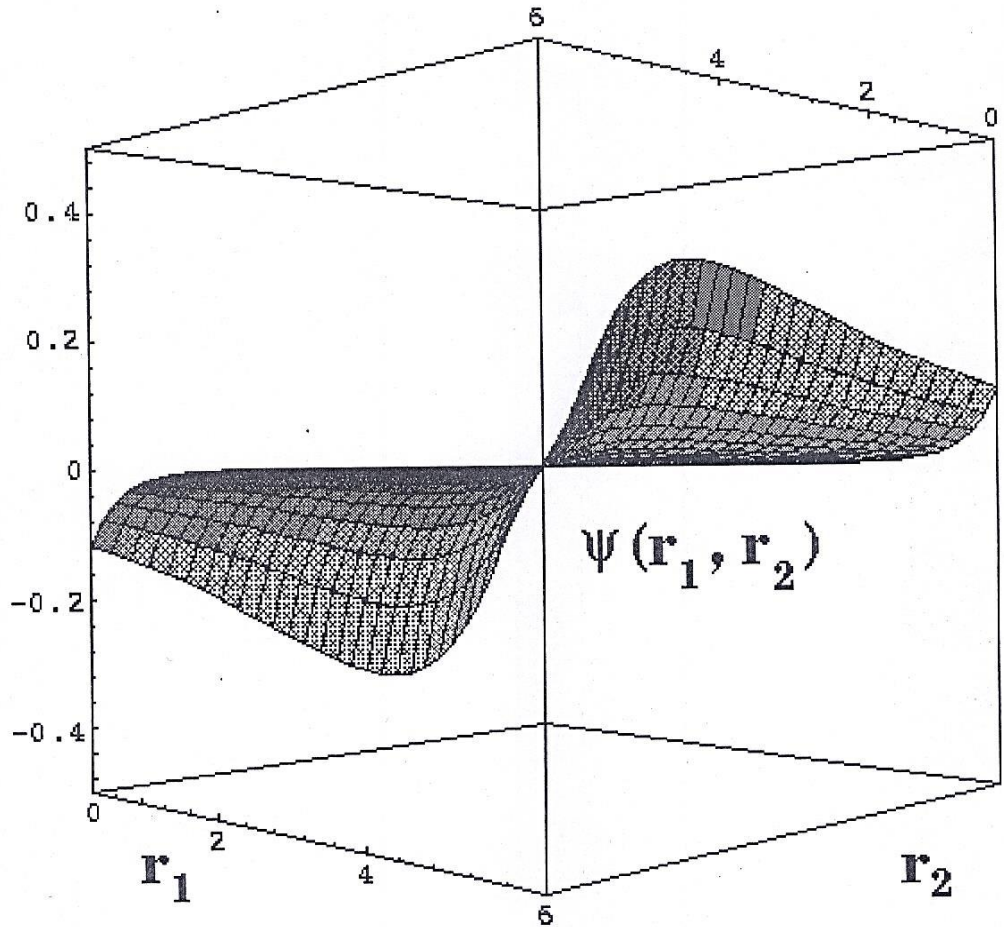

Figure 4  The plot of the wave function 5

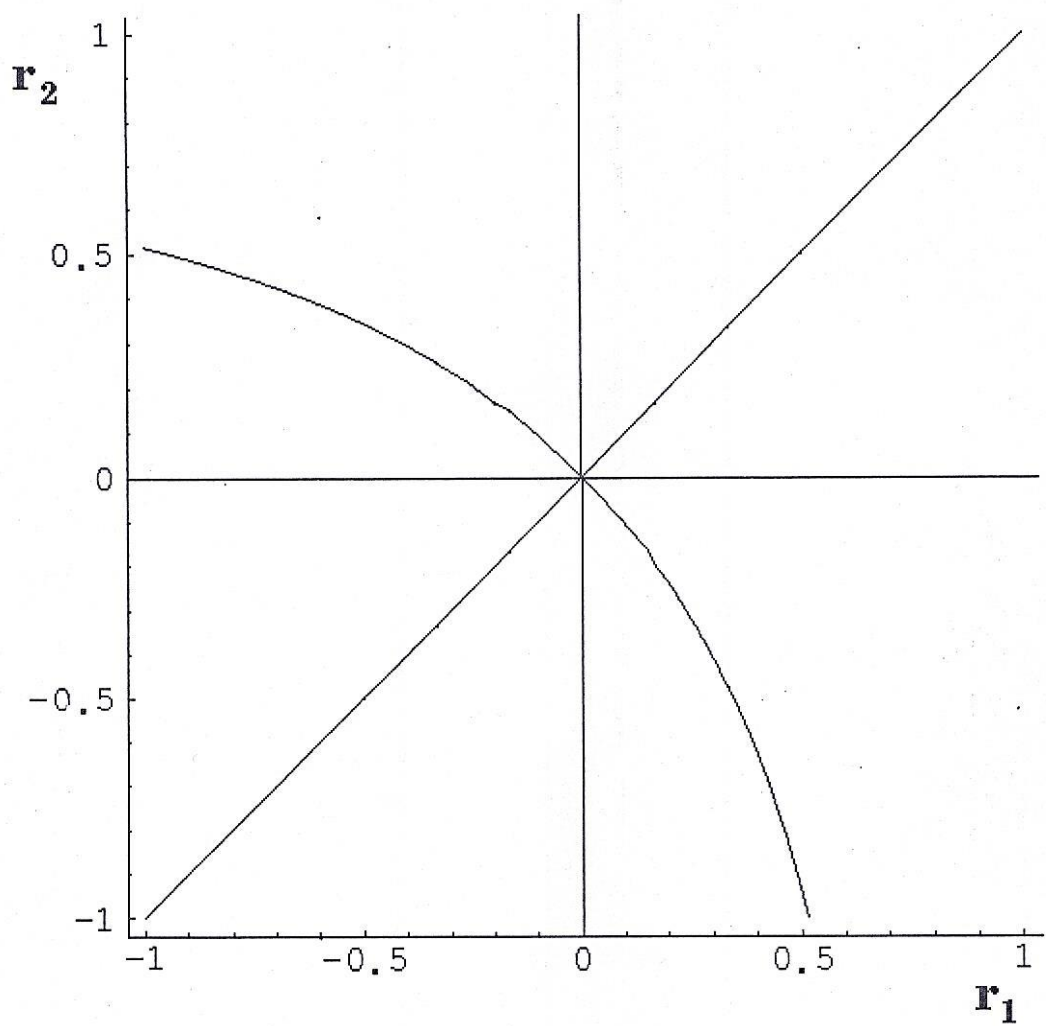

Figure 5  The plot of the nodal surface of the wave function 5

6 Configuration interaction (CI) wave function for the ground and first excited state of helium.

$\psi(\vec{r}_1, \vec{r}_2) = \sum_i c_i [\varphi_i(\vec{r}_1, \vec{r}_2) + \varphi_i(\vec{r}_2, \vec{r}_1)]$ where

$\varphi_i = e^{-\alpha_i r_1 - \beta_i r_2 - \sigma_i r_< - \tau_i r_> r_1^{a_i} r_2^{b_i} r_<^{s_i} r_>^{t_i}} \lambda_{l_{1i} l_{2i} LM}(\hat{r}_1, \hat{r}_2); i = 1,2$

$r_< = \min(r_1, r_2), r_> = \max(r_1, r_2)$

In general $\alpha_i = \beta_i = a_i = b_i = 0$

$s_i = 0,1,2............, t_i = 0,1,2.........$ if $s_i = 0,1$

$t_i = -s_i + 1, -s_i + 2.........$ if $s_i > 1$

In atomic physics, there exists a popular technique for including the effects of correlation in many electron systems. This technique is called configuration interaction(CI). In general, the wave functions can be represented as

$\psi = \sum_i c_i D_i$ where $D_i$'s are determinants each corresponding to a different orbital occupation scheme( i.e., different configurations). In this method, the energy is minimized as a function of mixing coefficients $c_i$ and hence known as (CI). The above wavefunction is a special case of CI wave functions for a particular choice of the following parameters:

a Ground state of Helium by CI. $\sigma_1 = 1.216604$, $\tau_1 = 1.920647$ $\sigma_2 = 1.994090$, $\tau_2 = 2.070513$, $c_1 = 77.457638$, $c_2 = -5.671781$, $a_1 = a_2 = b_1 = b_2 = 0$

$s_1 = t_1 = 0, s_2 = t_2 = 1$.

The above wave equation is used to calculate the energy for the ground state of helium. It does not have any nodes (as is required by the ground state wave functions). This wave function has the ability to deal with the electron-electron potential at $r_< = r_>$. Thus this possesses cusps as mentioned by Schwartz. Figure 6 shows the plot of Goldman's CI function for the ground state of He with our parameters.

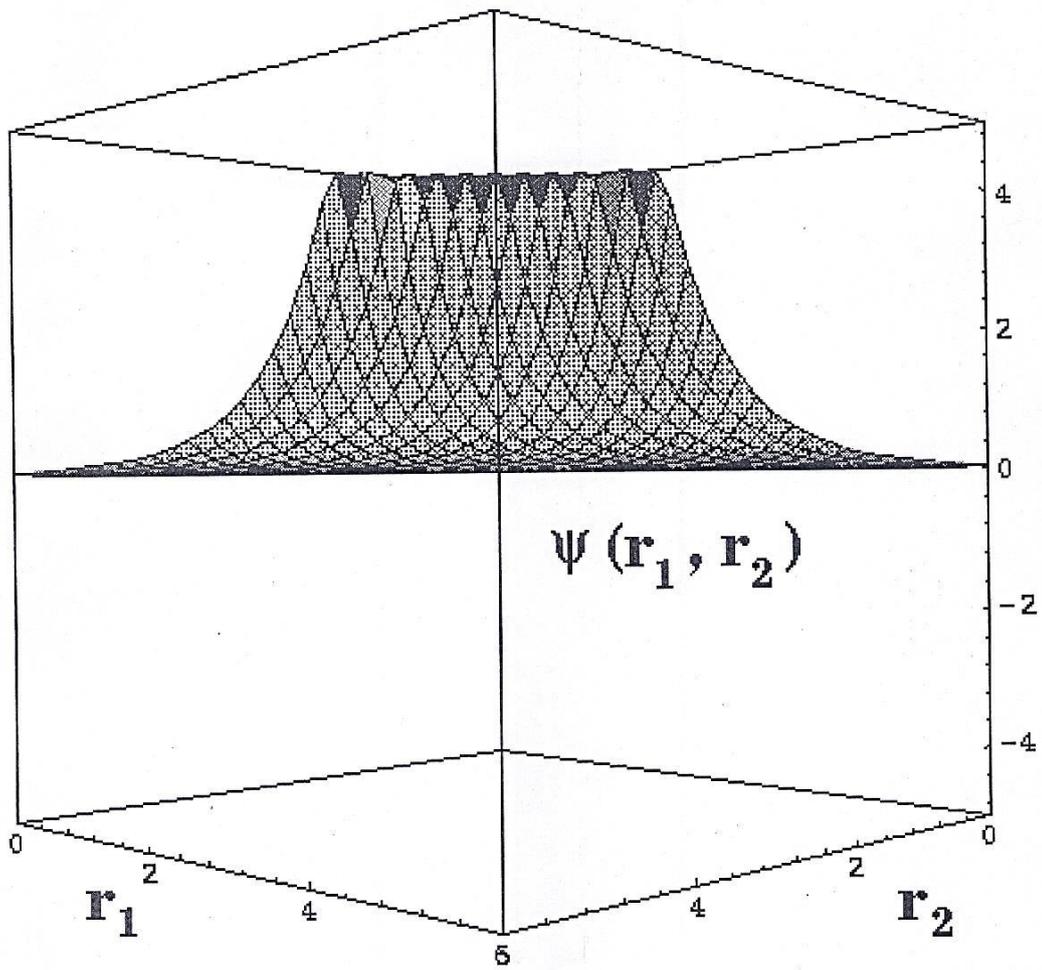

Figure 6 The plot of the wave function 6a

b First excited state of helium by CI:

$\sigma_1 = 1.981402$, $\tau_1 = 0.456199$, $\sigma_2 = 1.213401$ $\tau_2 = 1.810023$, $c_1 = 62454731$,

$c_2 = 13.154490$, $a_1 = a_2 = b_1 = b_2 = 0$

$s_1 = t_1 = s_2 = t_2 = 0$

The wave function in b is once again used to calculate the energy for the first excited state of helium. It satisfies the required nodal condition $r_1 = r_2$. Figure 7 shows the plot of the Goldman's function for the triplet S state with parameters in our optimization. Figure 8 shows the nodal surface of the Goldman's function 6b.

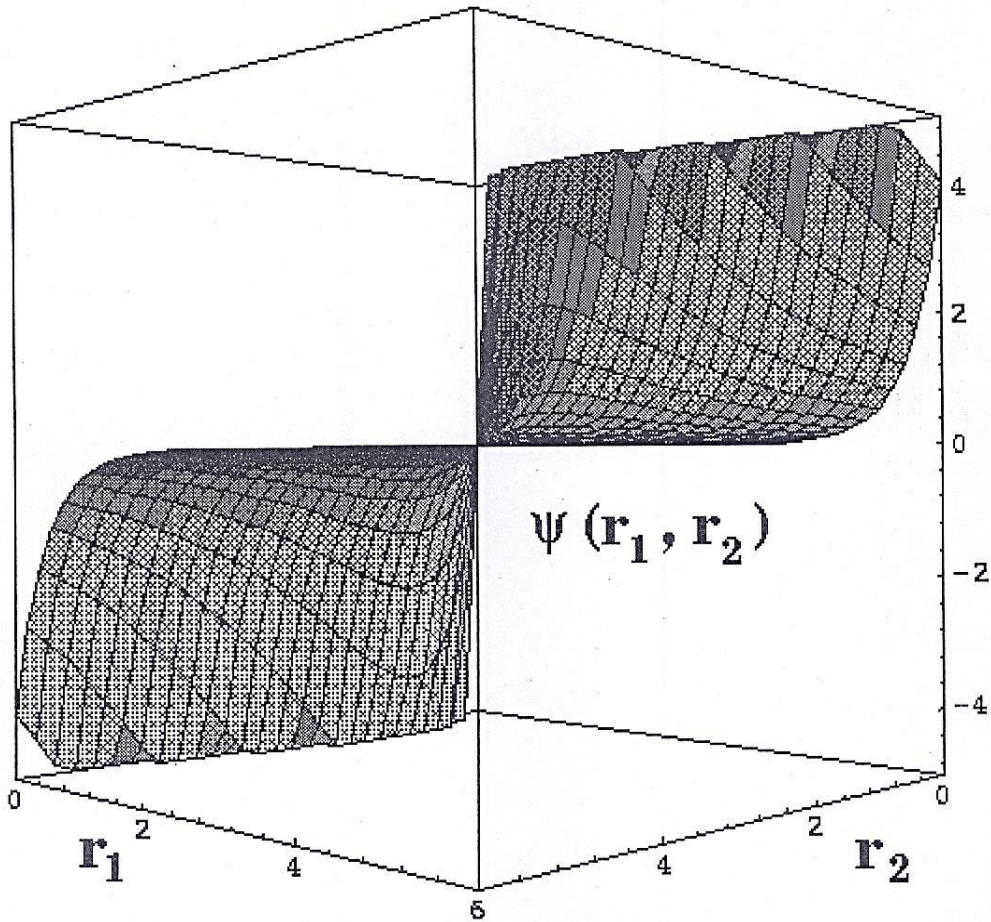

Figure 7 The plot of the wave function 6b

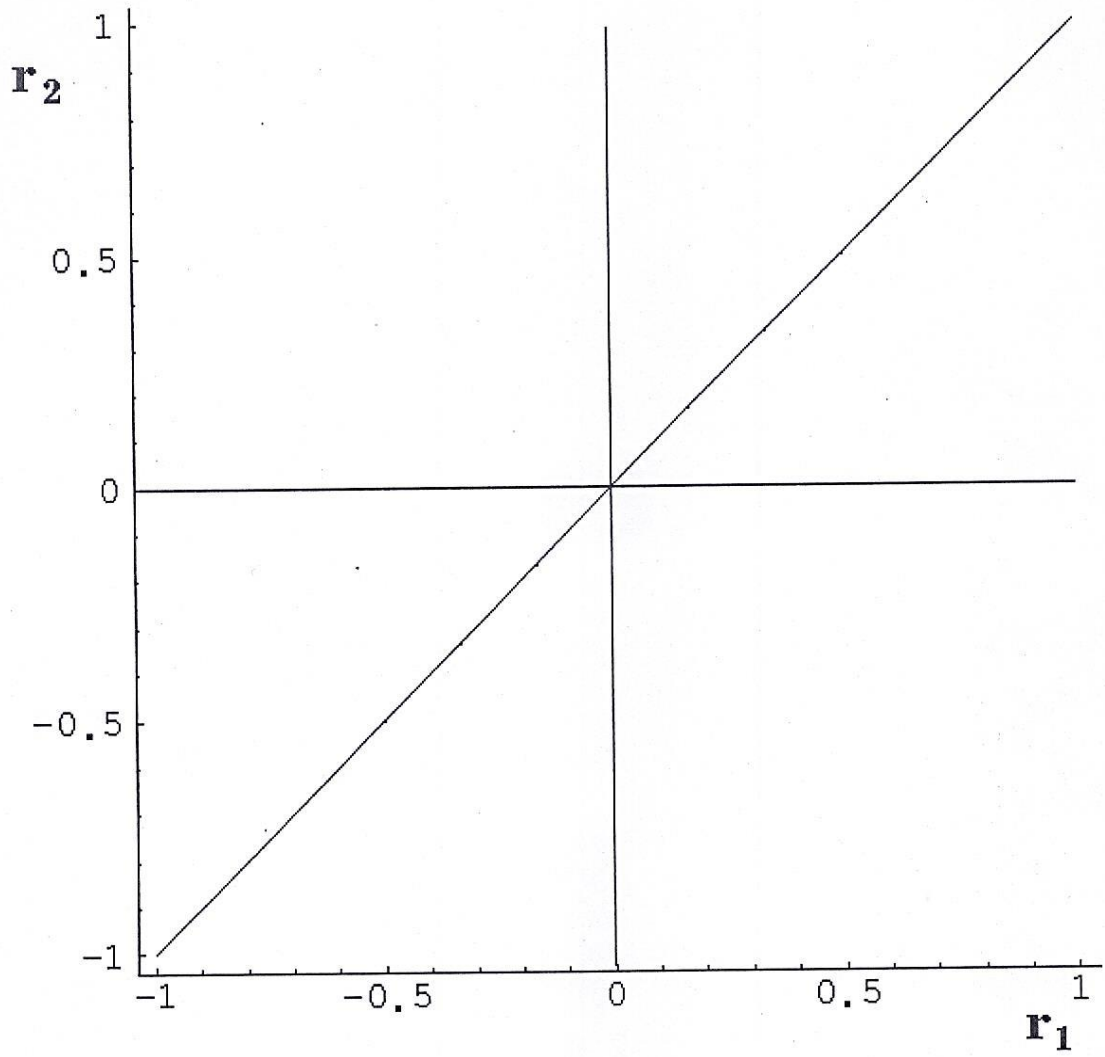

Figure 8 The plot of nodal surface of the wave function 6b

7 Wave function for $P_z$ state of Helium.

$$\psi(r_1, r_2, \vartheta_1, \vartheta_2) = (r_1 \cos\vartheta_1 + r_2 \cos\vartheta_2)[\exp(-\alpha_1 r_1 - \alpha_2 r_2) + \exp(-\alpha_2 r_1 - \alpha_1 r_2)].$$

It satisfies one of the two necessary nodal conditions for $P_z$ state, $z_1 = -z_2$ (proved in [6])

Table 1
Numerical results and statistical data for the first excited state of He with trial function.

3. $\psi(r_1, r_2) = (r_0 - r_1)e^{-\alpha_1 r_1 - \alpha_2 r_2} - (r_0 - r_2)e^{-\alpha_2 r_1 - \alpha_1 r_2}$ with $r_0 = 1, \alpha_1 = 1$ and $\alpha_2 = 2$

Scale =30,# of paths=1000K

| t  | zt       | ln(zt)   | ln(zt)/t | $\sigma$  | ln(zt)/t (ls fit) |
|----|----------|----------|----------|-----------|-------------------|
| 8  | 1.161759 | 0.149935 | 0.018741 | 0.000065  | 0.016554          |
| 16 | 1.739000 | 0.553310 | 0.034582 | 0.000071  | 0.034576          |
| 24 | 2.645342 | 0.972800 | 0.040533 | 0.000095  | 0.040583          |
| 32 | 4.240278 | 1.444628 | 0.045144 | 0.000373  | 0.043587          |
| 40 | 6.028673 | 1.796526 | 0.044913 | 0.000207  | 0.045389          |
| 48 | 9.628634 | 2.264741 | 0.047182 | 0.000369  | 0.046590          |

$$\lambda_0^{(0)} = -2.12412661 \quad \lambda_1 = -2.17536239(9)$$

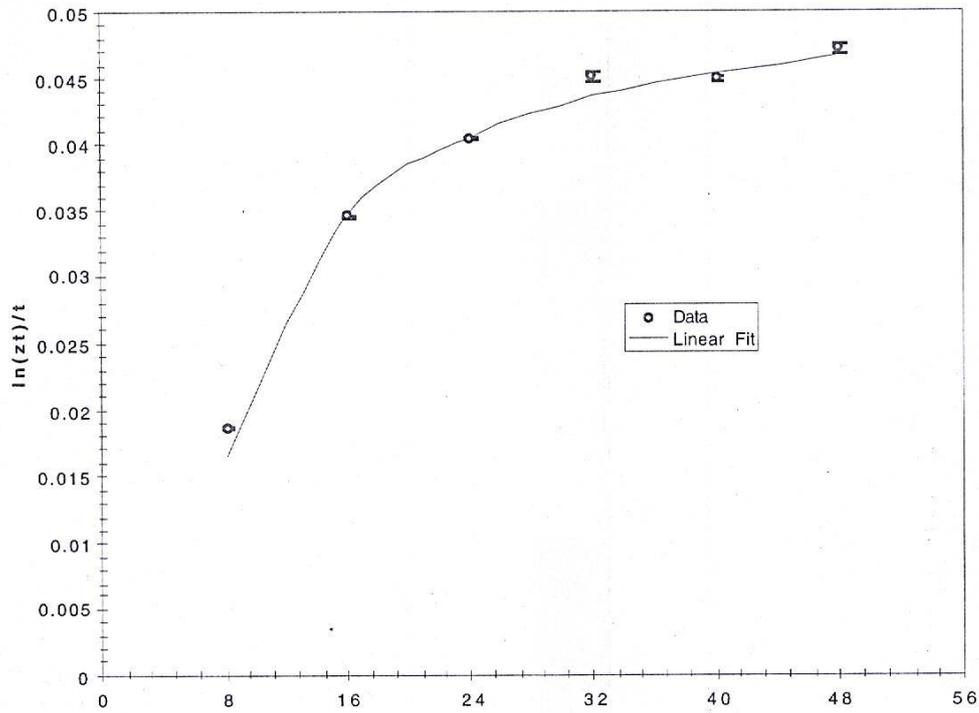

Figure 9 The plot of ln(zt)/t vs t for the wave function 3

Table 2
Numerical results and statistical data for the first excited state of He with trial function
5. $\psi(r_1, r_2) = (r_0 - r_1)e^{-\alpha_1 r_1 - \alpha_2 r_2} - (r_0 - r_2)e^{-\alpha_2 r_1 - \alpha_1 r_2}$ with $r_0 = 0.73351723, \alpha_1 = 0.636748$

Scale =30,# of paths=850K

| t | zt | ln(zt) | ln(zt)/t | $\sigma$ | ln(zt)/t (ls fit) |
|---|---|---|---|---|---|
| 8 | 0.958445 | -0.042443 | -0.005305 | 0.000022 | -0.005269 |
| 16 | 0.966878 | -0.033682 | -0.002105 | 0.000014 | -0.002124 |
| 24 | 0.974146 | -0.026194 | -0.001091 | 0.000011 | -0.001079 |
| 32 | 0.981974 | -0.018190 | -0.000568 | 0.000009 | -0.000556 |
| 40 | 0.989975 | -0.010075 | -0.000251 | 0.000008 | -0.000243 |
| 48 | 0.999053 | -0.000947 | -0.000019 | 0.000007 | -0.000034 |

$\lambda_0^{(0)} = -2.1742305 \; \lambda_1 = -2.1752508(1)$

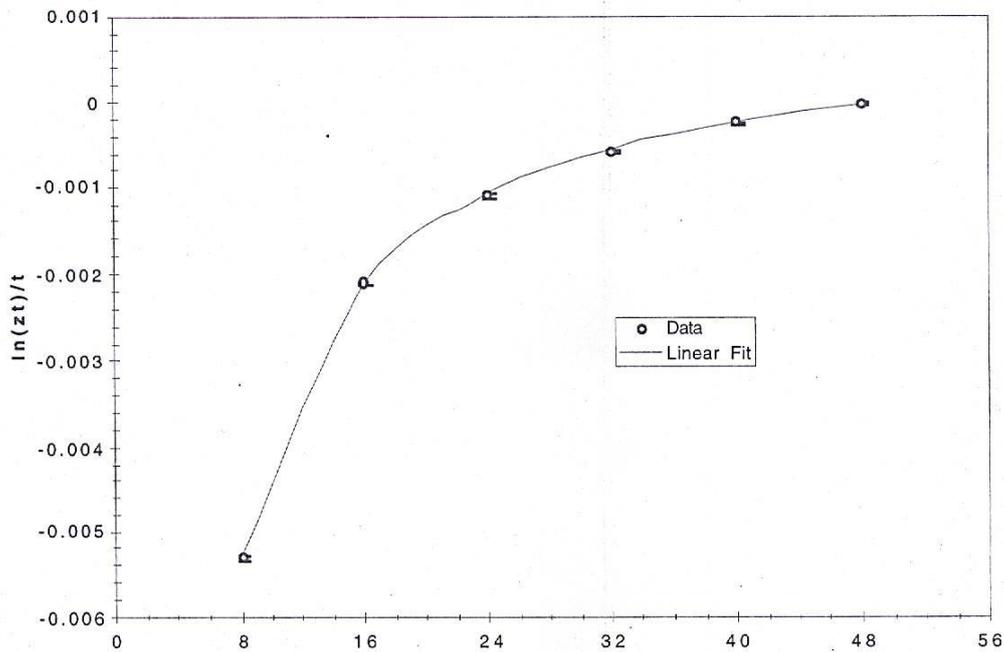

Figure 10 The plot of ln(zt)/t vs t for the wave function 5

Table 3
Numerical results and statistical data for the first excited state of He with trial function 6a
$\sigma_1 = 1.216604$, $\tau_1 = 1.920647$ $\sigma_2 = 1.994090$, $\tau_2 = 2.070513$, $c_1 = 77.457638$
$c_2 = -5.671781$, $a_1 = a_2 = b_1 = b_2 = 0$ $s_1 = t_1 = 0, s_2 = t_2 = 1$.

Scale =30,# of paths=150K

| t | zt | ln(zt) | ln(zt)/t | $\sigma$ | ln(zt)/t (ls fit) |
|---|---|---|---|---|---|
| 8 | 0.622624 | -0.473812 | -0.0059226 | 0.000243 | -0.060734 |
| 16 | 0.744703 | -0.294769 | -0.018423 | 0.000212 | -0.018527 |
| 24 | 0.893976 | -0.112076 | -0.004669 | 0.000214 | -0.004458 |
| 32 | 1.085791 | 0.082308 | 0.002572 | 0.000226 | 0.002575 |
| 40 | 1.312896 | 0.272223 | 0.006805 | 0.000232 | 0.006796 |
| 48 | 1.596764 | 0.467979 | 0.009749 | 0.000251 | 0.009610 |

$\lambda_0^{(0)} = -2.87651930$ $\lambda_1 = -2.9001983(27)$

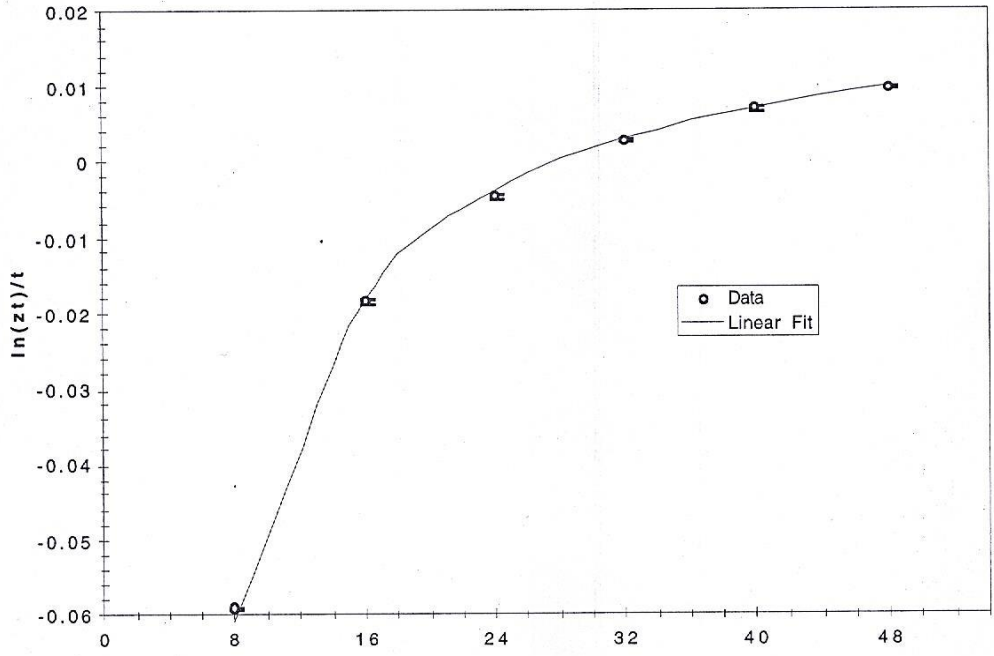

Figure 11 The plot of ln(zt)/t vs t for the wave function 6a

Table 4
Numerical results and statistical data for the first excited state of He with trial function
6b. $\sigma_1 = 1.981402$, $\tau_1 = 0.456199$, $\sigma_2 = 1.213401$ $\tau_2 = 1.810023$, $c_1 = 62454731$
$c_2 = 13.154490$, $a_1 = a_2 = b_1 = b_2 = 0$
$s_1 = t_1 = s_2 = t_2 = 0$

| | | Scale =30,# of paths=130K | | | |
|---|---|---|---|---|---|
| t | zt | ln(zt) | ln(zt)/t | $\sigma$ | ln(zt)/t (ls fit) |
| 8 | 1.091206 | 0.087283 | 0.010910 | 0.000087 | 0.012878 |
| 16 | 1.121974 | 0.115089 | 0.007193 | 0.000084 | 0.007307 |
| 24 | 1.144242 | 0.134742 | 0.005614 | 0.000078 | 0.005450 |
| 32 | 1.158441 | 0.147075 | 0.004596 | 0.000073 | 0.004522 |
| 40 | 1.170140 | 0.157123 | 0.003928 | 0.000070 | 0.003964 |
| 48 | 1.183700 | 0.168645 | 0.003513 | 0.000068 | 0.003593 |

$\lambda_0^{(0)} = -2.17401258$  $\lambda_1 = -2.17574917(8)$

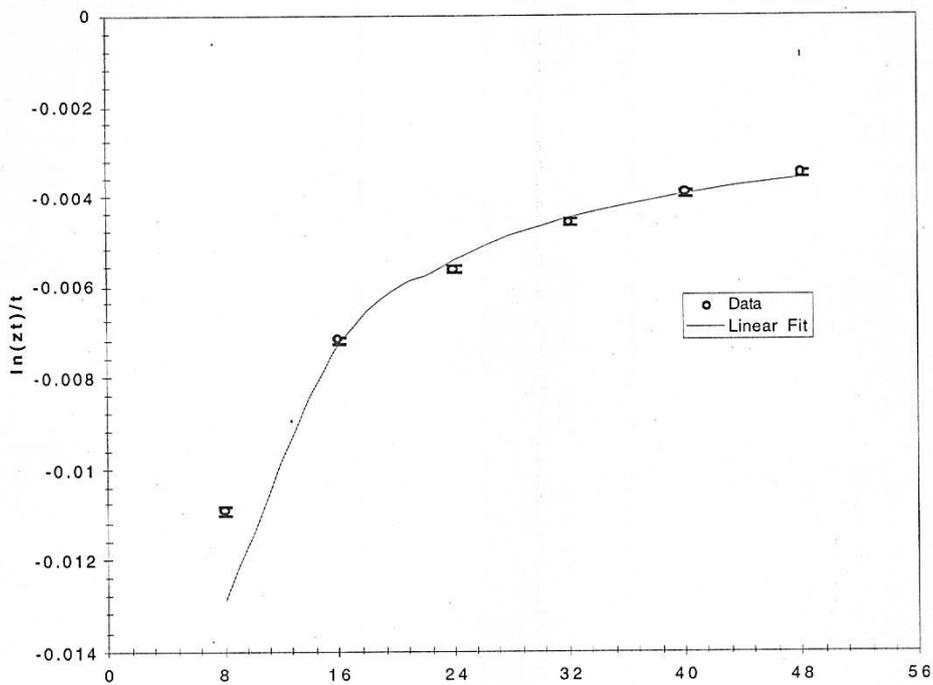

Figure 12  The plot of ln(zt)/t vs t for the wave function 6b

Table 5
Numerical results and statistical data for the first excited state of He with trial function 5.
$$\psi(r_1,r_2,\theta_1,\theta_2) = (r_1\cos\theta_1 + r_2\cos\theta_2)[\exp(-\alpha_1 r_1 - \alpha_2 r_2) + \exp(-\alpha_2 r_1 - \alpha_1 r_2)]$$

| | | Scale =30,# of paths=600K | | | |
|---|---|---|---|---|---|
| t | zt | ln(zt) | ln(zt)/t | $\sigma$ | ln(zt)/t (ls fit) |
| 8 | 0.482733 | -0.728291 | -0.091036 | 0.000194 | -0.090174 |
| 16 | 0.802182 | -0.220419 | -0.013776 | 0.000382 | -0.014894 |
| 24 | 1.231027 | 0.207848 | 0.008660 | 0.000299 | 0.010199 |
| 32 | 2.333108 | 0.847201 | 0.026475 | 0.001216 | 0.022745 |
| 40 | 4.734821 | 1.554943 | 0.03873 | 0.001610 | 0.030273 |
| 48 | 6.348487 | 1.848216 | 0.038504 | 0.000930 | 0.035292 |

$\lambda_0^{(0)} = -2.06460746 \quad \lambda_1 = -2.1250716(1)$

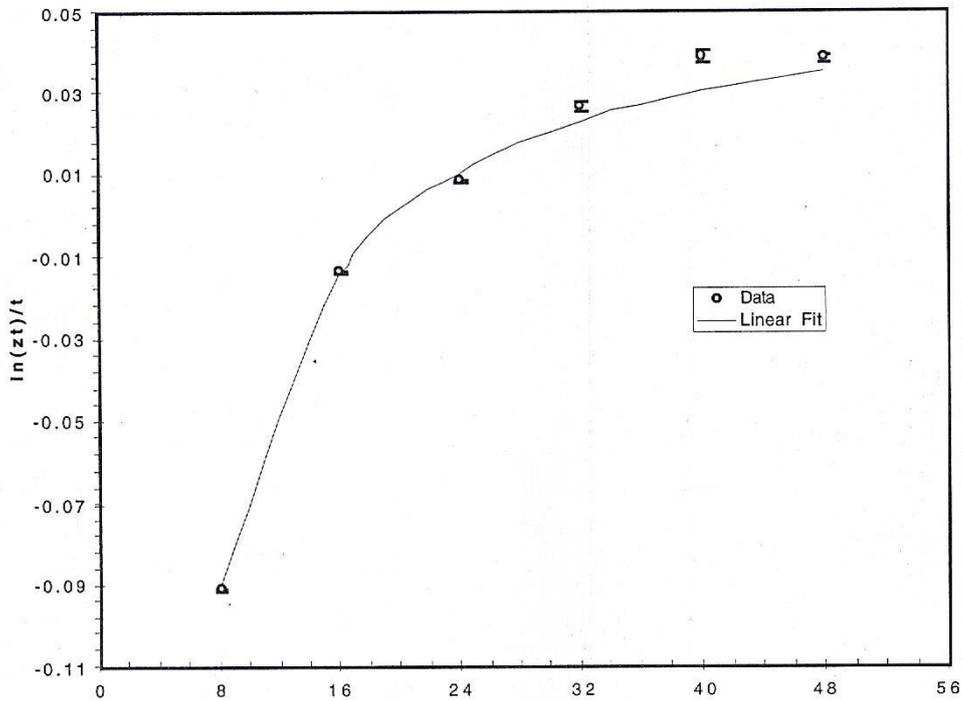

Figure 13  The plot of ln(zt)/t vs t for the wave function 7

Table 6. Atomic energies in Hartrees. Numbers in parenthesis are statistical errors. VQMC and Var are the results of elaborate Variational Monte Carlo and conventional Variational calculations respectively.

| | He($^1s_0$) | He($^3s_1$) | He($^1P_1$) |
|---|---|---|---|
| Var [a] | -2.903 724 377 034 119 5 | -2.175 229 378 236 791 30 | -2.123 843 086 498 093 |
| VQMC [b] | -2.903 724 372(5) | -2.175 229 376(3) | -2.123 843 89(7) |
| CI [c] | -2.876 651 930 | -2.174 012 58 | |
| GFK1 [d] | -2.903 726 82(6) | | |
| GFK1 [e] | -2.903 726 69(2) | | |
| GFK2 [f] | | -2.175 234 442(150) | |
| GFK2 [g] | | -2.175 229 14(237) | |
| GFK3 [h] | | -2.175 362 39(1) | |
| GFK5 [i] | | -2.175 250 8(1) | |
| GFK6 [j] | -2.900 019 83(27) | | |
| GFK6 [k] | | -2.175 749 17(8) | |
| GFK7 [l] | | | 2.125 076 16(1) |

[a] Ref[7]   [b] Ref[8]   [c] Refe[9]; our optimization]

[d] lin fit, this work using $\varphi_T^G$ in Ref[8]   [e] nlin fit, this work using $\varphi_T^G$ of Ref[8]

[f] lin fit, this work using $\varphi_T^E$ in Ref[8]   [g] nlin fit, this work using $\varphi_T^E$ in Ref[8]

[h] this work using $\varphi_T^E$ in Ref[5]   [i] this work; our optimization using $\varphi_T^E$ in Ref[5]

[j] this work; our optimization using $\varphi_T^G$ in Ref[9]   [k] this work; our optimization using $\varphi_T^E$ in Ref[9]

[l] this work using $\varphi_T^E$ in Ref[6]

It is evident from the numbers given in the Table 6(except for the Goldman's function for the ground state of He), that the GFK procedure implemented in our way gives the most accurate energy values known to date. Trial function 3 works satisfies the cusp condition and better converges than function 5 which does not have the cusp condition. For our choices of Goldman's parameter we see that for the triplet S state, GFK can improve the trial energy substantially, whereas GFK does not improve the ground state trial energy much. For the $P_z$ state, we use a trial function which obeys the required necessary conditions and GFK can also significantly improve the trial energy. Evidence obtained in these calculations convince the author that the GFK procedure can provide very accurate results for the lowest energy corresponding to a particular symmetry of many body systems, as opposed to Variational methods which can provide only the upper bound to the energy. This numerical procedure itself has a straightforward implementation to the Schrödinger Equation. To calculate energy we approximate an exact solution(i.e., the GFK representation) to the Schrödinger Equation, whereas most of the other numerical procedures approximate a solution to an approximate Schrödinger Equation. Approximations are involved only in the implementation of the GFK algorithm. Moreover this procedure is based on a well-defined mathematical formalism which makes the calculation rigorous even from the numerical perspective. This can provide exact results for the problems where the nodal conditions are exactly known.

Even the most elaborate Variational calculations give incorrect results for the following reasons. A typical Variational function is supposed to have the right cusp behaviour. The trial functions may not always be good enough to satisfy these conditions. Even if these trial functions satisfy all the necessary conditions, they could be largely restricted to the particular analytic forms rather than mimicking the true wave function of the system. In fact they might fail to provide the right energy of the system for not having enough Variational freedoms required to minimize the energy. The Generalized Feynman-Kac method can be applied to more complex systems for which energies are known up to a few significant figures from variational calculations and accuracy can be increased to include more significant figures.

Acknowledgements: The authors wish to thank Dr Steve A Alexander of Southwestern University for allowing them to use some of his trial functions.